\documentclass[a4paper,reqno]{amsart}

\long\def\drop#1{}

\let\limits\relax
\def\modd#1{\left|#1\right|}

\usepackage{graphicx}
\usepackage{geompsfi}

\def\pref#1{(\ref{#1})}

\newcommand{\R}{{\mathbb R}}
          
\newcommand{\pijl}{\longrightarrow}

\newcommand{\bdy}{\partial}             
         
\newcommand{\rn}[1]{\mathbb{R}^{#1}}
\newcommand{\caplet}[1]{\caption{{\small {\it {\textsf {#1}}}}}}

\newcommand{\Ao}{\mathcal A^1_{r_0}}
\newcommand{\At}{\mathcal A^2_{r_0}}
\newcommand{\te}{\theta}

\newcommand{\df}{\begin{description}\item[Definition]}

\newtheorem{thm}{Theorem}[section]
\newtheorem{lem}[thm]{Lemma}
\newtheorem{cor}[thm]{Corollary}
\newtheorem{defn}[thm]{Definition}
\makeatletter
\let\c@remark\c@thm
\let\p@remark\p@thm
\makeatother
\def\theremark{\thesection.\arabic{remark}}
\newenvironment{remark}%
    {\par\medbreak\refstepcounter{remark}%
         {\noindent\textbf{Remark~\theremark.\ }}}%
    {\par\medbreak}

\newcommand{\la}{\lambda}

\newcommand{\ta}{\theta}

\newcommand{\Om}{\omega}

\begin{document}

\title[Link and Writhe for Open Rods]%
{A consistent treatment of link and writhe for open rods,
and their relation to end rotation}

\thanks{Submitted to Archive for Rational Mechanics and Analysis}
\author[Van der Heijden]{Gert H.~M.~van der Heijden\dag\email{\dag ucesgvd@ucl.ac.uk, 
\ddag peletier@cwi.nl, \S rplanque@cwi.nl.}}  
\address[\dag]{Centre for Nonlinear Dynamics, University College London, Gower Street, London WC1E 6BT, UK.}
\author[Peletier]{Mark A.~Peletier\ddag}
\address[\ddag,\S]{Centrum voor Wiskunde en Informatica, Kruislaan 413, 1098 SJ Amsterdam, The Netherlands.} 
\author[Planqu\'e]{Robert Planqu\'e\S}

\begin{abstract}
We combine and extend the work of Alexander \& Antman
\cite{alexander.82} and Fuller \cite{fuller.71,fuller.78} to give a
framework within which precise definitions can be given of topological
and geometrical quantities characterising the contortion of open
rods undergoing large deformations under end loading. We use these
definitions to examine the extension of known results for closed rods to
open rods. In particular, we formulate the analogue of the celebrated
formula $Lk=Tw+Wr$ (link equals twist plus writhe) for open rods and propose
an end rotation, through which the applied end moment does work, in the form
of an integral over the length of the rod. The results serve to promote the
variational analysis of boundary-value problems for rods undergoing large
deformations.
\end{abstract}

\maketitle

\pagestyle{headings}

\section{Introduction}

In a variational analysis of an elastic structure that is
acted upon by end forces and moments one needs to consider the work done
by the applied loads. The work done by the applied force does not usually
present any problems. One requires the distance travelled by the force,
which is usually easy to obtain. More problems occur in determining the
work done by the applied moment. Here one requires the end rotation as
`seen' by the moment, which may lead to complications if large deformations
are allowed. This paper discusses the ambiguities associated with
this end rotation and shows how a consistent treatment can be obtained.

If the structure is very long it may be modelled by an infinitely long rod.
This has the advantage that powerful techniques from dynamical systems
(viewing arclength along the rod as time) \cite{champneys.99,heijden.00.1}
and variational analysis \cite{peletier.01} can be used. The natural class
of solutions to consider in this case are localised solutions, which decay
sufficiently rapidly towards the ends (other solutions would have infinite
strain energy and would therefore be non-physical). The boundary conditions
to be imposed on such solutions are simple: end tangents are aligned and the
ends of the rod do not interfere with the {\em localised} deformation. The
work done by the torsional load is then simply the product of applied end
twisting moment $M$ and relative end rotation $R$ (the angle that has gone
into one end of the rod in order to produce the deformation, starting from
a straight and untwisted reference configuration and keeping the other
end fixed).

However, any real-world problem (be it a drill string, a marine pipeline or
a DNA supercoil) deals with a finite-length rod. For such a rod more
complicated boundary conditions may be encountered. For instance, the end
tangents need not be aligned, so that an end rotation that could be used
in an energy discussion is not straightforward to define. But even in the
case of aligned end tangents complications arise if large deformations are
allowed. These complications are the subject of this paper.

\begin{figure}[t]
\centerline{\psfig{figure=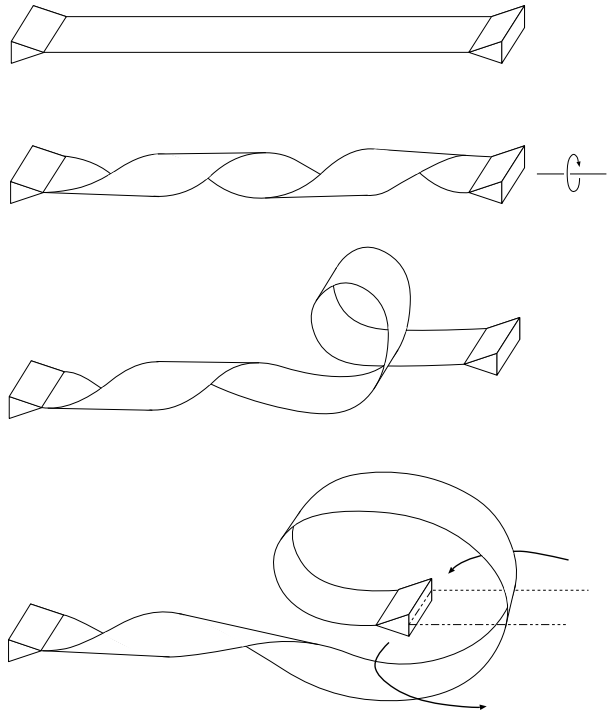,width=8cm}}
\caption{Four rod configurations that are not distinguished by a prescribed
process (homotopy) of boundary conditions from a fixed reference
configuration: one can go from (a) to (b) either through an end
rotation or through `looping'. (After~\cite{alexander.82}).}
\label{fig:strip}
\end{figure}

Let us demonstrate the issue with an example. If we rotate the right end of the
rod in Fig.~\ref{fig:strip}(a) through an angle of $-4\pi$ we obtain the
configuration shown in (b). If we now move the right end of the rod to the
left, the rod pops into a looped configuration as shown in (c). If we
move the right end further and make the loop pass around the right clamp,
as in (d), and then pull the ends out again, we return to configuration (a)
without having rotated the ends. (Note that the process illustrated in
Fig.~\ref{fig:strip}(d) requires that whatever supports the right clamp must
be released to allow the passage of the rod behind it.) We conclude that we
can go from configuration (a) to configuration (b) either by end rotation or
by `looping'. What, then, is the `real' end rotation of the deformation
(a) $\to$ (b) `seen' by the applied moment and therefore pertinent to an
energy analysis?

It is apparent from this ambiguity that it is not sufficient,
\emph{a priori}, to consider only the initial and final configurations of
a deformation process to decide on the end rotation. We need to be told
{\em how} the rod was deformed from the one into the other, i.e., we need
to know the deformation {\em history}, not just locally of the ends but
globally of the entire rod. (This path dependence suggests a relation with
a geometric phase; details on this are found in
\cite{hannay.98,maggs.01,starostin.02}.)

Alexander \& Antman \cite{alexander.82} address the end rotation ambiguity
for fixed boundary conditions by imagining the open rod to be part of a
closed rod. A natural restriction to a class of open rod deformations is
then obtained by demanding that the closed continuation should not undergo
self-intersections. Applied to Fig.~\ref{fig:strip} this would mean that
throwing the rod around the clamp as in (d) is excluded because it requires
an intersection with the closure (indicated by the dotted lines). In the
restricted class of deformations the configurations (b) and (c) can then be
assigned a unique end rotation of $R=-4\pi$ relative to (a). The ambiguity
in the end rotation is thus resolved.

To distinguish between different classes of deformations Alexander \& Antman
use the \emph{link} of the closed rod, i.e., the topological linking number
of two lines drawn on opposite sides of the unstressed rod. Since two
unknotted closed rods with equal linking number may be deformed into each
other without undergoing self-intersections, there is a one-to-one
correspondence between the classes of admissible deformations with unique $R$
and the values of the link. The link is thus seen to be related to the end
rotation. (Note that the $-4\pi$ versus 0 ambiguity in the end rotation
illustrated in Fig.~\ref{fig:strip} is reflected in the jump in link by
$-2$ as the closed rod intersects itself.)

The connection between link for closed rods and end rotation for open rods
has been discussed by many authors. The level of detail varies, but
one usually argues from the link of a closed rod to the link of an open rod,
which is then identified with the end rotation. Often this involves the
introduction of a closure and the use of the celebrated formula
\cite{calugareanu.61,fuller.71,white.69}
\begin{equation}
Lk=Tw+Wr,
\label{eq:ltw}
\end{equation}
which expresses the link $Lk$ (of a closed rod) in terms of the twist $Tw$
(a local property, in the sense that it can be found by integrating a density
along the length of the rod) and the writhe $Wr$. This writhe, which is only
a function of the centerline of the rod, is not a local property but several
expressions exist for the calculation of the writhe of an arbitrary closed
curve \cite{aldinger.95}. Some of these expressions make sense for open
curves as well and therefore suggest an extension of writhe from closed to
open curves \cite{starostin}. The final result, then, is a formula for the
open link which in terms of a suitable Euler-angle representation takes the
simple form
\begin{equation}
\label{eq:link}
\text{open link} = \frac{1}{2\pi} \int [\dot \phi + \dot \psi].
\end{equation}

This formula cannot be expected to hold true in general since it would make
link a local property, which it is not. Indeed, the generalisation of writhe
from closed to open curves used in obtaining \pref{eq:link} is subject to a
geometrical condition (see Section~\ref{sec:results}). 
Certainly general validity of
(\ref{eq:link}) is prevented by the polar singularity inevitably associated
with Euler angles. 
In Section~\ref{sec:Euler} formula (\ref{eq:link}) will
be derived within a limited class of deformations.

\medskip

The present paper improves on previous results in the following ways.
\begin{enumerate}
\item The closure introduced by Alexander \& Antman is effective
in the definition of admissible deformations if the supports
are fixed in both angle and position. This is adequate for a large
class of experimental situations. If, however, the supports
are allowed to move, then it is not possible to use one and the same
closure for all deformations: different deformations require
different closures. Thus the desired distinction between classes of
deformations is lost (one can always move the closure along, so that
it does not `get in the way' during the deformation).
%
%
We resolve the issue by limiting the class of admissible closures
in such a way that the separation into different classes of deformations,
characterised by the link, is preserved under arbitrary movement of the
supports.

\item We define a precise class of deformations within which link, twist, and
writhe are well-defined. We carefully examine the restrictions imposed on
the deformations and show in which sense they are necessary. 
As mentioned above, 
for a consistent definition of link and writhe it is necessary to work within a
class of \emph{homotopies of rods}, which connect a given open
rod to a reference configuration. Despite the requirement of such a
connection, the newly introduced 
writhe and link themselves are independent of the
choice of connection: within the class the
writhe and link only depend on the given open rod and the reference 
configuration.

\item We show that the link, twist and writhe defined for open rods
satisfy the classical equality \pref{eq:ltw}. Furthermore, within the class
of admissible deformations the link is given by \pref{eq:link}. In the
special case that the end tangents of the open rod are aligned the link
coincides with the end rotation. Our results thus formalise current practice
in the literature based on \pref{eq:link}. However, while
most applications of formula~\pref{eq:link} can be shown to fall into our
class of deformations, recent experiments in molecular biology do not always
do so and care is required in energy discussions (we discuss this in
Section 4). 
Indeed, we would claim that usage of \pref{eq:link} presupposes
the framework that we here discuss, and consequently is subject to the
limitations that we describe.

\end{enumerate}

The organisation of the paper is as follows. In Section~\ref{sec:results}
we first define our class of open rods. For the elements of this class we
define link, twist and writhe and derive the extension of \pref{eq:ltw} to
open rods. In Section~\ref{sec:Euler} we independently define the end
rotation and show it to be equal to the open link. We also derive
\pref{eq:link}. In Section~\ref{sec:crit} we critically review the defining
conditions of the class of rods considered, illustrating their 
relevance with counterexamples. Section~\ref{sec:disc} discusses our
work in the light of previous work in the literature. 

\section{Results: Link, Twist and Writhe}
\label{sec:results}

For our purposes a rod is a member of the
set
\begin{multline*}
\mathcal{A}^0 = \bigl\{\, (r,d_1) \in C^2([0,L]; \rn{3} \times S^2)
\text{ such that }
|\dot{r}| \ne 0, 
\;\dot{r} \cdot d_1 = 0,\\
\text{and $r$ is non-self-intersecting } \bigr\}.
\end{multline*}
Here and in the following an overdot denotes
differentiation with respect to the spatial variable $s$. The curve~$r$ 
is thought of as the
centerline of a physical rod (of length $L$) and $d_1(s)$ as a material
vector in the section at $s$. As alternatives to `rod' the terms `ribbon'
\cite{fuller.78,aldinger.95} and `strip' \cite{alexander.82} are also
used. A closed rod is an element of~$\mathcal A^0$ for which begin and end 
connect smoothly.

To each point on the centerline of the rod we can attach an
orthonormal right-handed frame $(d_1(s), d_2(s), d_3(s))$
of directors by setting 
\begin{equation}
\label{def:deriv_d123}
d_3(s) = \dot{r}(s)/|\dot{r}(s)|\qquad\text{and}\qquad
d_2(s) = d_3(s) \times d_1(s).
\end{equation}
These directors track the varying
orientation of the cross-section of the rod along the length of the rod.
The twist of a closed rod $(r,d_1)$ is now defined by
\begin{eqnarray}
\label{eq:twist_def}
Tw(r,d_1) := \frac1{2\pi}
\oint\limits_{r} \dot{d}_1(s) \cdot d_2(s)\, ds.
\end{eqnarray}
It measures the number of times $d_1$ revolves around $d_3$ in the
direction of $d_2$ as we go around the rod.

Let $r_1$ and $r_2$ be two non-intersecting curves. Then the link of
$r_1$ and $r_2$ is defined by
\begin{eqnarray}
\label{eq:lk_2_curves}
Lk(r_1, r_2) := \frac1{4\pi}\oint\limits_{r_1}\! \oint\limits_{r_2} 
\frac{[\dot{r}_1(s) \times \dot{r}_2(t)] \cdot [r_1(s) - r_2(t)]}
{|r_1(s) - r_2(t)|^3}\,ds dt.
\end{eqnarray}

The writhe of a closed curve $r$ is
\begin{eqnarray}
\label{eq:writhe_def}
Wr(r) := \frac1{4\pi} \oint\limits_{r}\! \oint\limits_{r} 
\frac{[\dot{r}(s) \times \dot{r}(t)] \cdot [r(s) - r(t)]}
{|r(s) - r(t)|^3}\,ds dt.
\end{eqnarray}
The argument of this integral is the pullback of
the area form on $S^2$ under the Gauss map $\rn{2} \pijl S^2$, 
\[
G : (r(s), r(t)) \longmapsto \frac{r(s) - r(t)}{|r(s) - r(t)|},
\]
so that the writhe may be interpreted as the signed area on $S^2$ 
that is covered by
this map. 
For each direction $p \in S^2$ the
signed multiplicity of the Gauss map (i.e., the number
of points $(s,t)$ for which $G(s,t) = p$, weighted by the
sign of $p\cdot [G_s\times G_t]$) equals the {\it directional writhing number},
the number of signed crossings of the
projection of $r$ onto a plane orthogonal to the vector 
$p$~\cite{fuller.71,aldinger.95}. 
In other words,
the writhe of a closed curve is equal to the directional
writhing number averaged over all directions of $S^2$.

The link, twist and writhe of a closed rod are related by the
well-known C\u alug\u areanu-White-Fuller Theorem
\cite{calugareanu.61,white.69,fuller.71}:
\begin{thm} 
\label{thm:lk=tw+wr}
Let $(r,d_1) \in \mathcal{A}^0$ be a closed rod as defined above. Then
\begin{eqnarray}
Lk(r,d_1) = Tw(r,d_1) + Wr(r).
\end{eqnarray}
\end{thm}

We review two classical theorems by Fuller which are of interest to us.
Note that at each point $r(s)$, the unit tangent
$t(s) = \dot{r}(s)/|\dot{r}(s)|$ traces out a closed curve on $S^2$,
called the tantrix. Fuller's first theorem relates the writhe of the
curve to the area $A$ enclosed by the tantrix on $S^2$:
\begin{eqnarray}
\label{eq:fuller1}
Wr(r) = \frac{A}{2\pi} - 1 \ (\mathrm{modd} \ 2).
\end{eqnarray}
Note that the equality modulo two is necessary since the area enclosed by
a curve on $S^2$ is only defined modulo $4\pi$.

\begin{figure}
\centerline{\psfig{figure=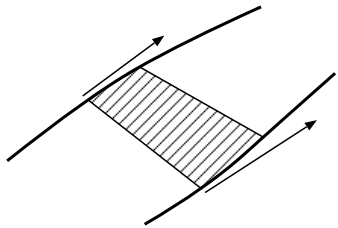,height=3cm}}
\caption{The argument of the integral in~\pref{eq:fuller2} is the
area swept out by the geodesic connecting the curves $t_1$ and $t_2$ on $S^2$.}
\label{fig:FullerII}
\end{figure}

The second theorem, stated in detail as Theorem~\ref{thm:fuller} below,
gives under certain conditions
a formula for the difference in writhe between
two closed curves $r_1$ and $r_2$ that can be continuously deformed
into each another (see Figure~\ref{fig:FullerII}):
\begin{eqnarray}
\label{eq:fuller2}
Wr(r_1) - Wr(r_2) = \frac 1{2\pi} \int \frac{t_2 \times t_1}{1 + t_1
\cdot t_2} \cdot (\dot{t}_1 + \dot{t}_2).
\end{eqnarray}

\medskip

We now proceed with the introduction of the
set for which the open
link, twist and writhe will be defined. 
For the definition of this set we choose a closed planar reference curve 
$r_0\in C^2([0,M];\R^3)$ for some $M>L$.
\begin{defn}
\label{def:C^1}
\begin{alignat*}2
\Ao = \Big\{\ &\ \quad&& (r,d_1) \in \mathcal{A}^0 \ 
\text{such that }
\exists (\bar{r}, \bar{d}_1) \in
        C^2([0,M] \times [0,1]; \rn{3} \times S^2):
\\
&1.&& \text{for each } \la,\ \bar{r}(\cdot,\la) 
  \text{ is an unknotted, non-self-intersecting}\\
&&&\quad \text{ closed curve}, 
  \\
&2.&& \dot{\bar{r}}(s,\la)\cdot\bar{d}_1(s,\la) = 0 \quad\text{for }s \in [0,M],\
\la \in \{0,1\}, \\
&3.&& (\bar{r}, \bar{d}_1) (s,1) = (r,d_1)(s)  \quad \text{for }s \in [0,L], \\
%
%
&4.&& \bar r(s,0) = r_0(s) \quad  \text{for }s \in [0,L], \\ 
&5.&& \dot{\bar{r}}(s,0) \cdot \dot{\bar{r}} (s, \la) > -1\quad
         \text{for }s \in [0,M],\ \la \in [0,1], 
\global\def\nonoppcond{5} \\
&6.&& \{\bar{r}(s,\la)\colon s\in[L,M]\} \text{ is a planar curve for } 
  \la = 0 \text{ and } \la=1 ,\\ 
&&&\quad \text{ and these two planes are parallel }
\global\def\planaritycond{6}
\Big\}.
\end{alignat*}
\end{defn}
$\Ao$ can be thought of as a class of open rods $(r,d_1)$
that can be connected by a homotopy---satisfying certain requirements---%
to the reference curve $r_0$.
The part of the closed rod parametrised by $s\in[L,M]$ is called the
closure.

Some of the conditions above are more straightforward than others.
Parts 1 and~2 state that $(\bar r, \bar d_1)$ is a homotopy of
well-behaved closed rods, and by parts~3 and~4 the homotopy contains
the original open rod $(r,d_1)$ at $\lambda=1$, and
the reference curve at $\lambda=0$. Parts
\nonoppcond\ and~\planaritycond\ contain the essential elements of this
definition. Part~\nonoppcond\ is the same non-opposition condition that appears
in the statement of Fuller's theorem~(Theorem~\ref{thm:fuller}) and is
required for the conversion of the writhe to a single-integral expression.
Part~\planaritycond, 
which states that the closure should be planar at the beginning
and the end of the homotopy, is central in the construction. These last
two conditions are discussed more fully in Section~\ref{sec:crit}.

Note that curves $r_0$ exist for which $\Ao$ is empty: if the three vectors
$\dot r_0(0)$, $\dot r_0(L)$, and $r_0(L)-r_0(0)$ are independent, then
the open curve $r_0$ can not be closed by a planar closure, so that
the set of homotopies with planar closures that connect to $r_0$ is empty.
\bigskip

We are now in a position to define the new functionals open link,
open twist and open writhe for open rods.

\begin{defn} 
\label{def:LTW}
Let $(r,d_1)$ be a rod in $\Ao$. Then the open twist of
$(r,d_1)$ is 
\begin{multline*}
Tw^o(r,d_1) := \frac1{2\pi}\int\limits_0^L \dot{d}_1 \cdot (\dot{r} \times d_1)
\,ds,\\
\end{multline*}
the open writhe of $r$ is
\begin{multline*}
Wr^o(r) := Wr(\bar{r}(\cdot,1)) = \frac 1{4\pi} \int\limits_0^M\! \int\limits_0^M
\frac{[\bar{r}(s,1) - \bar{r}(t,1)]\cdot[\dot{\bar{r}}(s,1) \times
\dot{\bar{r}}(t,1)]} {|\bar{r}(s,1) - \bar{r}(t,1)|^3}\, ds dt, \\
\end{multline*}
and the open link of $(r,d_1)$ is 
\begin{multline*}
Lk^o(r,d_1) := Lk(\bar{r}(\cdot,1),\bar{d}_1(\cdot,1)) - 
\frac{1}{2\pi}\int_L^M \dot{\bar{d}}_1(s,1) \cdot (\dot{\bar{r}}(s,1) \times
\dot{\bar{d}}_1(s,1))\, ds.\\
\end{multline*}
\end{defn}
Note that in the last definition we subtract any twist the closure might
have. It follows directly from the construction
that the new concepts also satisfy the classical relationship:
\begin{cor}
\label{cor:lk0=tw0+wr0}
Let $(r,d_1) \in \Ao$ be an open rod. Then 
\[
Lk^o(r,d_1) = Tw^o(r,d_1) + Wr^o(r).
\]
\end{cor}

\begin{thm}
For any open rod $(r, d_1) \in \Ao$, the open twist, writhe,
and link are well-defined.
\label{thm:well-def}
\end{thm}
For writhe and link this is non-trivial, as different
homotopy closures  $(\bar r, \bar d_1)$ might be expected to give rise
to different values.

\begin{proof}
We first state Fuller's second theorem in a more precise form.
\begin{thm}
\label{thm:fuller}
Let $r_\lambda$ $(0\leq\lambda\leq1)$ be a homotopy of closed non-self-intersecting
curves, regularly parametrized with a common
parameter $s\in[0,L]$. Let $t_\lambda$ be the tantrix of $r_\lambda$. 
If ${t}_0(s) \cdot {t}_{\lambda}(s) > -1$ for 
all $s \in [0,L], \la \in [0,1]$, 
then 
\begin{eqnarray}
Wr(r_1) - Wr(r_0) = \frac1{2\pi} \int\limits_0^L 
\frac{t_0 \times t_1}{1 + t_0 \cdot t_1}
\cdot \big( \dot{t}_0 + \dot{t}_1\big) ds.
\end{eqnarray}
\end{thm}
To our knowledge, a rigorous proof was first given by
Aldinger {\it et al.}~\cite{aldinger.95}. 

\medskip

To prove Theorem~\ref{thm:well-def} for the open writhe, 
let $(\bar{r}, \bar{d}_1)$ be a homotopy associated to $(r, d_1)$. By
definition, $\bar{r}(\cdot,0)$ and $\bar{r}(\cdot,1)$ are planar for $s 
\in [L,M]$. Let us denote the planes by $V_0$ and $V_1$; these are parallel
by Definition~\ref{def:C^1}.\planaritycond. 
Let $V$ be the plane through the origin
parallel to both. 

We have defined the class of open rods  $\Ao$
such that Theorem \ref{thm:fuller} can be applied. Denote the
tantrices of $\bar{r}(\cdot,0)$ and $\bar{r}(\cdot,1)$ by $t_0$ and $t_1$
respectively. Then
\begin{eqnarray}
\label{eq:ful-wr}
Wr(\bar{r}(\cdot,1)) - Wr(\bar{r}(\cdot,0)) &\!\!=\!\!& 
\frac1{2\pi}\int\limits_0^M \frac{t_0(s) \times
t_1(s)}  
{1 + t_0(s)\cdot t_1(s)} \cdot
(\dot{t}_0(s) + \dot{t}_1(s))\,ds. 
\end{eqnarray}
The argument of the integral vanishes for $s\in [L,M]$:
since $t_0(s)$, $t_1(s) \in V$ for $s \in
[L,M]$ we have 
$t_0(s) \times t_1(s) \perp V$ and
$\dot{t}_0(s) + \dot{t}_1(s) \in V$ for $s \in [L,M]$. Hence 
\[
[t_0(s) \times t_1(s)] \cdot [\dot{t}_0(s) +
\dot{t}_1(s)] = 0.
\]
Moreover, since $\bar{r}(\cdot,0)$ is planar, $Wr(\bar{r}(\cdot,0)) = 0$. 
We conclude
\begin{eqnarray}
\label{eq:int-var-r}
Wr(\bar{r}(\cdot,1)) &=& 
\frac1{2\pi}\int\limits_0^L \frac{t_0(s) \times
t_1(s)} 
{1 + t_0(s)\cdot t_1(s)} \cdot
(\dot{t}_0(s) + \dot{t}_1(s))\,ds.
\end{eqnarray}
Since this integral only depends on the reference curve and the
open rod itself, and is otherwise independent of the choice of closure and
homotopy, this proves the claim for the writhe.

\medskip
For the link, let $(\bar{r}, \bar{d}_1)$ be a homotopy associated to $(r, d_1)$ by
Definition~\ref{def:C^1}. Denote the closed curves $\bar{r}(\cdot,1)$ and
$\bar{d}_1(\cdot,1)$ by $\hat{r}$ and $\hat{d}_1$ respectively.
We denote the open twist evaluated over an
interval $s \in [a,b]$ by $Tw^o_{[a,b]}$. Then
\begin{eqnarray}
Tw^o_{[L,M]}(\hat{r},\hat{d}_1) &=&
Tw^o_{[0,M]}(\hat{r},\hat{d}_1) - 
Tw^o_{[0,L]}(\hat{r}, \hat{d}_1) \nonumber \\
&=& Tw^o_{[0,M]}(\hat{r},\hat{d}_1) - 
Tw^o(r,d_1). \nonumber
\end{eqnarray}
Hence
\begin{eqnarray}
Lk^o(r,d_1) &=& Lk(\hat{r},\hat{d}_1) - Tw^o_{[L,M]}(\hat{r},\hat{d}_1)
\nonumber \\
        &=& Lk(\hat{r},\hat{d}_1) -
Tw(\hat{r},\hat{d}_1) + Tw^o(r,d_1)  
\nonumber \\
        &=& Wr(\hat{r}) + Tw^o(r,d_1) \nonumber \\
        &=& Wr^o(r) + Tw^o(r,d_1)\label{eq:lk0=wr+tw0}.
\end{eqnarray}
It follows that $Lk^o(r,d_1)$ is independent of the chosen
closure $(\bar r, \bar d_1)$.
\end{proof}

Equation~\pref{eq:int-var-r} is an important motivation of this
work, since it expresses the writhe in terms of a single rather than a
double integral. For the purpose of variational analysis this is an
obvious advantage. It is especially useful when the link, and therefore
indirectly the writhe, can be identified with the rotation of the
ends; this requires that the end tangents remain equal throughout the 
deformation, and this case is treated in the next section.

\drop{
Its usefulness is limited, however, when the tangents at the ends of the
rod do not remain aligned throughout the deformation. This is due to 
two facts: First, a scalar end rotation can only be defined in the case
of constant end tangents (see Section~\ref{sec:Euler} below). Secondly,
the integral in~\pref{eq:single-int-const} may depend on $t_0$; only when the 
end tangents are aligned does this dependence disappear, as demonstrated
by the following lemma.

}

\begin{remark}
For simplicity we have chosen to introduce one class $\Ao$
as the basis for the definitions of open link, twist and writhe.
For each of the three definitions separately, however, not all of
Definition~\ref{def:C^1} is required. Open twist can be defined
directly in terms of $(r,d_1)$, without the need for an extension;
open link requires a  closure, but no homotopy; and open writhe
requires a homotopy, but the director~$d_1$ can be disposed of.
\end{remark}

\begin{remark}
\label{rem:cont_conn}
A natural question to ask is whether the open link, twist and writhe
reduce to their classical counterparts when an open rod is transformed
into a closed rod by lining up and connecting the ends.
This is not the case, as we demonstrate in Section~\ref{sec:crit}.
\end{remark}

\section{Results: End-rotation and Euler angles}
\label{sec:Euler}
It is common in applications to assume that the end tangents of the
buckled rod are kept constant and equal during the deformation process.
For comparison with an end rotation we introduce this additional
condition. Throughout this section we also assume that $r_0|_{[0,L]}$ is
straight; without loss of generality we assume that
$\dot r_0$ is a constant unit vector $v$ on $[0,L]$.
Finally, again without loss of generality we choose the director $d_1$ constant
on the reference curve $r_0|_{[0,L]}$:
\begin{defn}
\begin{align*}
\At =  \bigl\{\,(r,d_1) \in \Ao\colon\;
&\dot r_0(s) = v \in S^2 \text{ for all } s\in [0,L],\\ 
&\dot{\bar{r}}(0,\la) =
\dot{\bar{r}}(L,\la) = \dot{\bar{r}}(M,\la) 
= v \text{ for all }  \la \in [0,1], \\
&\bar d_1(s,0) = \bar d_1(0,0) \text{ for all }s\in[0,L]\;\bigr\}.
\end{align*}
\end{defn}
%
The following formula is a direct consequence of Theorem~\ref{thm:fuller}:
\begin{cor}
\label{cor:wr=single-int}
Let $(r,d_1) \in \At$ and let $t$ be the tantrix 
of $r$. Then
\begin{eqnarray}
\label{eq:single-int-const}
Wr^o(r) = \frac 1{2\pi} \int\limits_0^L 
 \frac{v \times t(s)}{1 + v \cdot t(s)} 
    \cdot \dot{t}(s) \, ds.
\end{eqnarray}
\end{cor}

In the present case of a straight $r_0|_{[0,L]}$ the dependence 
on $r_0$ of the open writhe of a given open rod 
takes a particularly simple form:

\begin{thm}
\label{thm:wr-rot-invariance2}
Under the conditions of Corollary~\ref{cor:wr=single-int},
let $\Omega = S^2\setminus \{-t(s): s\in [0,L]\}$. Then the function
\[
v \in S^2 \longmapsto \frac1{2\pi}\int_0^L \frac{v\times t(s)}{1+v\cdot t(s)}
  \cdot \dot t(s) \, ds
\]
is constant on connected components of\/ $\Omega$.
\end{thm}
The proof is given in the appendix.
The interpretation of this theorem is as follows: when the end tangents are 
aligned, the tantrix given by the rod (without closure) forms
a closed curve on~$S^2$. The integral above represents 
`area enclosed by the curve' for a given `choice of area' ({\em cf.}
\pref{eq:fuller1}). When the
vector $v$ crosses the set $\{-t(s): s\in[0,L]\}$ the geodesic connections
between $v$ and $t(s)$ change direction, causing the integral to
represent a different choice of area, and therefore causing the integral
to jump by $4\pi$.

\medskip

With fixed end tangents we can
introduce a fourth quantity, the end rotation.
We denote $\bdy(\cdot)/\bdy \la$ by~$\bdy_{\la}(\cdot)$.
\begin{defn}
\label{def:Rlambda}
Let $(r, d_1) \in \At$, and let 
$\bar d_3(\cdot,\cdot) = \dot{\bar r}(\cdot,\cdot)/\modd{\dot{\bar r}(\cdot,\cdot)}$,
$\bar d_2 = \bar d_3\times \bar d_1$. We define the 
end rotation by
\[
R(r,d_1) :=
\int\limits_0^1 \bdy_{\la}\bar{d}_1(L,\la) \cdot 
        \bar d_2(L, \la)\,d{\la} 
-\int\limits_0^1 \bdy_{\la}\bar{d}_1(0,\la) \cdot 
        \bar d_2(0, \la)\,d{\la}.
\]
\end{defn}
To study the relationship between end rotation and open link, twist and
writhe we introduce a particular choice of 
Euler angles for an open rod $(r,d_1)$. Recall that for
every $s \in [0,L]$ there is an orthonormal director frame
$(d_1(s), d_2(s), d_3(s))$. We express this frame in terms of
angles $\ta, \psi, \phi$ with respect to a fixed basis $(e_1, e_2,
e_3)$ as follows 
\begin{align}
\label{eq:d123}
d_1 &= (-\sin\psi\sin\phi + \cos\psi\cos\phi\cos\ta)\,e_1 + \nonumber\\
&\phantom{=\ }(\cos\psi\sin\phi + \sin\psi\cos\phi\cos\ta)\,e_2 - 
\cos\phi\sin\ta\,e_3, \nonumber \\
d_2 &= (-\sin\psi\sin\phi - \cos\psi\sin\phi\cos\ta)\,e_1 + \\
&\phantom{=\ }(\cos\psi\cos\phi + \sin\psi\sin\phi\cos\ta)\,e_2 - 
\sin\phi\sin\ta\,e_3, \nonumber \\
d_3 &= \cos\psi\sin\ta\,e_1 + \sin\psi\sin\ta\,e_2 + \cos\ta\, e_3. \nonumber
\end{align}
This choice of Euler angles follows Love \cite[art.~253]{love.44}.
For rods in the class $\At$ we choose $e_3$ parallel
to $v$; note that by this choice the 
non-opposition condition~\nonoppcond\ in Definition~\ref{def:C^1}
coincides with avoidance of the
Euler-angle singularity at $\te=\pi$. Therefore the smoothness assumptions
on $(\bar r, \bar d_1)$ in $\At$ imply $C^1$-regularity
for $\phi$, $\psi$, and $\te$.

\begin{lem}
\label{pr:Rla=phi+psi}
Let $(r,d_1) \in \At$ be an open rod with an associated
homotopy  $(\bar{r},\bar{d}_1)$,
and let $\bar d_2$ and $\bar d_3$ be constructed from
$\bar r$ and $\bar d_1$ according to~\pref{def:deriv_d123}. 
Let $\phi, \psi, \te:[0,M]\times [0,1]\to\R$ be
the Euler-angle representation of $(\bar{d}_1,\bar d_2,\bar d_3)$.
Then 
\[
R(r,d_1) 
 = \int\limits_0^L [\dot{\phi}(s,1) + \dot{\psi}(s,1)]\, ds.
\]
\end{lem}
From this Lemma we conclude
\begin{cor}
\label{cor:R1-well-def}
For open rods $(r,d_1) \in
\At$, $R(r,d_1)$ is independent of the choice of extension
$(\bar r,\bar d_1)$.
\end{cor}
\drop{
Note that for a given reference \emph{curve} $r_0$,  $R(r,d_1)$ necessarily
depends on the choice of the reference \emph{rod},
i.e.\ on the choice of the director field $\bar d_1(\cdot,0)$ and therefore
on $\dot \phi(\cdot,0) + \dot \psi (\cdot,0)$.
}
%

\begin{proof}[Proof of Lemma \ref{pr:Rla=phi+psi}]
Using the definitions of the Euler angles, we find
\[
\bdy_{\la} \bar d_1(s,\la) \cdot \bar d_2(s,\la) = \bdy_\la{\phi}(s,\la) 
+ \bdy_\la{\psi}(s,\la) \,
\cos \ta(s,\la).
\]
For $s = 0,L$ and $\la \in [0,1]$ we have set 
$\bar d_3(s,\la) = e_3$, and hence
$\theta(s,\la) = 0$ for $s = 0,L$; therefore 
$\bdy_{\la} \bar d_1(s,\la) \cdot \bar d_2(s,\la) =$ $\bdy_\la{\phi}(s,\la) +
\bdy_\la{\psi}(s,\la)$ for $s = 0,L$. 


Since ${\phi}+ {\psi}$ is a continuously differentiable function on
$V:=[0,L]\times [0,1]$, the integral of the tangential 
derivative of $ \phi +\psi$ along $\bdy V$  vanishes:
\[
\oint_{\bdy V} \frac{\bdy}{\bdy\tau} (\phi+\psi) = 0,
\] 
where $\tau$ is the clockwise-pointing unit vector tangential to 
$\bdy V$.

Hence
\begin{align*}
\lefteqn{\int_0^L [\bdy_s{\phi}(s,1) + \bdy_s{\psi}(s,1)]\, ds 
  - \int_0^L [\bdy_s {\phi}(s,0) + \bdy_s{\psi}(s,0)]\, ds = }
\hskip1cm&\\
&=\int_0^1 [\bdy_\la{\phi}(L,\la) + \bdy_\la{\psi}(L,\la)]\, d\la - 
\int_0^1 [\bdy_\la{\phi}(0,\la) + \bdy_\la{\psi}(0,\la)]\, d\la \\
&=\int_0^1 [\bdy_\la{\phi}(s,\la) + \bdy_\la{\psi}(s,\la)]\, d\la\Big|_{s=0}^{s=L}  \\
&=\int_0^1 \bdy_{\la} \bar d_1(s,\la) \cdot \bar d_2(s,\la)\, d\la
\Big|_{s=0}^{s=L} = R(r,d_1). \nonumber
\end{align*}
\end{proof}

\bigskip
Within this framework Euler-angle formulae for open link, twist and
writhe are obtained:
\begin{lem}
\label{lem:wr-tw-Eangles}
Let $(r, d_1)$ be an open rod in $\At$. Then 
\begin{eqnarray}
\label{eq:wr0-Eangles}
Wr^o(r) = \frac{1}{2\pi}\int\limits_0^L \dot{\psi}(s) (1 - \cos \ta(s))\, ds,
\end{eqnarray}
and
\begin{eqnarray}
\label{tw0-Eangles}
Tw^o(r,d_1) = \frac{1}{2\pi}\int\limits_0^L [\dot{\phi}(s) +
\dot{\psi}(s) \cos \ta(s)]\,ds.
\end{eqnarray}
\end{lem}

\begin{proof}
The formula for twist is easily found by using (\ref{eq:d123})
in the definition of twist, as in the proof of Lemma
\ref{pr:Rla=phi+psi}.
For the writhe we apply Corollary~\ref{cor:wr=single-int} and use
the fact that $v= e_3$. 
\end{proof}
The main result of this section states that for rods in $\At$
end rotation is equal to the open link:

\begin{thm}
\label{thm:lk0=r=r1}
Let $(r,d_1)$ be an open rod in $\At$. Then
\begin{equation}
\label{eq:lk0=r=r1}
R(r,d_1) = 2\pi Lk^o(r,d_1) = \int_0^L [\dot{\phi}(s) + \dot{\psi}(s)]\, ds
=\phi(L)+\psi(L)-\phi(0)-\psi(0).
\end{equation}
\end{thm}

\begin{proof}
By Corollary \ref{cor:lk0=tw0+wr0} and Lemma \ref{lem:wr-tw-Eangles}
we obtain
\[
Lk^o(r,d_1) = Wr^o(r) + Tw^o(r,d_1) = \frac{1}{2\pi} \int_0^L[ \dot{\phi}(s) +
\dot{\psi}(s)]\, ds. 
\]
Since $\int_0^L [\dot{\phi}(s) + \dot{\psi}(s)]\, ds = R$
by Lemma \ref{pr:Rla=phi+psi} we have the desired result. 
\end{proof}

\section{Critique of the approach}
\label{sec:crit}

The example of Fig.~\ref{fig:strip} shows that end rotation can only be
defined for a deformation history. For the purpose of analysis of elastic
structures this dependence on deformation history is undesirable. The
approach in this paper, which is shared by many others (see the next section),
is therefore to construct a class of
deformation histories (homotopies) within which the end rotation {\em can}
be expressed in terms of the initial and final states only. In this section
we critically review the essential ingredients of this approach.

\textbf{The closure.} 
We obtain a separation into deformation classes by the introduction of a
closure. The classes are characterised by the link of the closed rod-closure
combination. The price we pay with this construction is the dependence on
the choice of closure, which at first sight might seem to be a defect of the
formulation. As Alexander \& Antman \cite{alexander.82} point out, however,
this dependence is entirely natural: the precise form of the closure can be
regarded as describing the way the rod is supported. Different systems of
supports necessarily allow different classes of deformations.

Although in most applications throwing the rod around the clamp as illustrated
in Fig.~\ref{fig:strip}(d) is physically prevented, there do exist
exceptions to this rule.
In some recent experiments
DNA molecules are manipulated with the help of
a magnetic bead attached to the end of the molecule and held in a magnetic
trap which allows the simultaneous application of a force and a moment
\cite{strick.96}. In this case no rigid mechanical support is present
and the molecule is free to loop around the beaded end. If this happens
repeatedly, then the end can rotate over an arbitrarily large angle under the
applied moment without a concomitant change in configuration
(\emph{cf}.~Fig.~\ref{fig:strip}: after a rotation of the end by $-4\pi$ the
rod returns to its original shape). Thus one might estimate the wrong energy,
by $\pm 4\pi M$ for each `looping', if a deformation would go outside the
class, i.e., if link was not conserved. However, Rossetto \& Maggs
\cite{rossetto.xx} in a recent paper show that for micron-sized beads the
applied tension in many experiments is large enough (on the order
of femtonewtons) to make these link
violations rare, and they proceed to introduce a closure to study
configurations of constant link.

Incidentally, the fact that a rotation of $\pm 4\pi$,
and not one of 
$\pm 2\pi$, brings one back to where one started has its origin in the
topological nature of SO(3), the group of rotations in
$\R^3$. Specifically, SO(3) is not simply connected: for every
rotation $R \in$ SO(3) there are two homotopy classes of paths from
the identity of the group to $R$. This means that for a given rod
orientation there are two distinct classes of configurations for the
rod which cannot be deformed into each other while keeping the ends
fixed. Rods with any even number of end turns (including zero) lie in
one class; rods with any odd number of turns lie in the other.
The same topological property forms the basis of the famous Dirac Belt Trick
\cite{kaufmann.91}, which in classrooms is often illustrated by rotating a
cup, held in the palm of one's hand, twice around a vertical axis by a
suitable motion of the arm to bring both cup (with contents) and arm
back to their initial positions (see \cite{feynman.87} for a demonstration).

\textbf{The reference curve.}
The class $\Ao$ is defined for a fixed closed reference curve $r_0$ 
(which may or may not include the unstressed centerline of the rod). The
open writhe $Wr^o$  
will, in general, depend on the choice of this curve; on the one hand,
by the fact that the reference curve restricts the class of
admissible homotopies via the non-opposition condition, and
on the other hand, by 
the explicit dependence on $t_0$ in~\pref{eq:int-var-r}.
Similarly, the end rotation $R$ will depend on $r_0$, as is to be expected
since $R$ is defined (in Definition~\ref{def:Rlambda}) 
as the end rotation incurred in
deforming $(r_0,\bar d_1(\cdot,0))|_{[0,L]}$ into $(r,d_1)$.

For certain cases, however, the dependence can be described more
precisely, as in Theorem~\ref{thm:wr-rot-invariance2} for straight
reference curves. 
Though beyond the scope of this paper, it is also
possible to prove a similar result under rotation of 
more general reference curves.
For instance, one could extend Definition~\ref{def:Rlambda}, 
Lemma~\ref{pr:Rla=phi+psi} and Corollary~\ref{cor:R1-well-def}
to a larger class than $A^2_{r_0}$ by requiring of $r_0|_{[0,L]}$ 
only that its end
tangents be equal. A complication would arise, however, in that violation
of the non-opposition condition would no longer coincide with the Euler-angle
singularity. Consequently, the non-opposition condition would no longer
assure us of $C^1$-regularity of $\phi$, $\psi$, and $\theta$.

\textbf{The non-opposition condition cannot be dropped.}
The non-op\-position condition listed in the definition of $\Ao$
(condition~\nonoppcond) is imposed by the application of
Theorem~\ref{thm:fuller}. This might seem to be merely a technical
restriction: after all, the open writhe is defined in terms of the writhe
of the closed curve (Definition~\ref{def:LTW}) and the latter is
well-defined even if, somewhere along the homotopy, the non-opposition
condition is violated. Therefore it might be expected that the statement of
well-posedness holds true without condition 6 (even though our proof
evidently does not), and that only non-self-intersection of the closed
structure is required.

\begin{figure}[ht]
\centering
\centerline{\psfig{figure=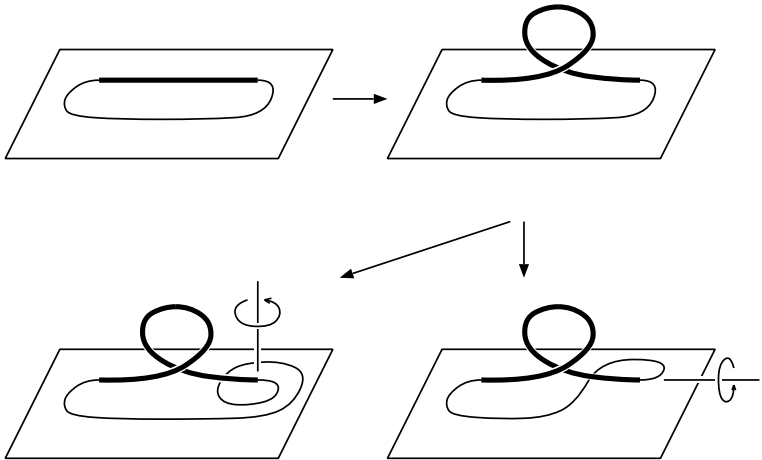,width=0.8\hsize}}
\caption{An example showing that the non-opposition condition~\nonoppcond\
in Definition~\ref{def:C^1} cannot be disposed of. The three final states
(b), (c) and (d), with identical shapes for the open rod, have different
values of writhe. In going from (b) to either (c) or (d) the non-opposition
condition is violated.}
\label{fig:count_ex_genWrithe}
\end{figure}

In fact the situation is not that simple. 
Figure~\ref{fig:count_ex_genWrithe}
shows homotopy paths connecting the reference configuration
(a) with the deformed rod-closure combinations (b), (c) and (d), where 
the rod itself (represented by the thick line) is the same in each of
the three deformed states. We can imagine the deformed rod to be nearly
planar, with the two strands crossing at a short distance from each other.
Then the rod and its closure in case (b) have writhe close to $-1$.%
\footnote{This may be verified by using the characterisation of writhe
as the average of the directional writhing number, as explained
in Section~\ref{sec:results}. This number is determined by
counting signed crossings in a projection of the curve onto a plane.}
In case (c) one adds or subtracts 1 to the writhe of the rod-closure
combination for each full turn of the end. The writhe of the combination
can therefore be made arbitrarily large. In case (d), finally, the writhe
is close to $-2$.

It is not difficult to see that one may construct a
homotopy between (a) and (b) without violating the
non-opposition condition, provided the loop
has been twisted through an angle strictly less than $\pi$. Since the
continuation homotopies to (c) and (d) satisfy all conditions of
Definition~\ref{def:C^1} other than the non-opposition condition, it
follows from Theorem~\ref{thm:well-def} that these homotopies cannot be
constructed without violating the non-opposition condition. This may also
be verified by inspection. 

This example shows that simply removing condition~\nonoppcond\ from
Definition~\ref{def:C^1} leads to ambiguities in the definition
of writhe (and therefore of link). The example also suggests that
if a well-defined writhe is to be constructed without the inclusion
of the non-opposition condition, then additional
restrictions must be imposed on the closure. In homotopy (c) the
closure remains planar throughout the homotopy, but the end tangents
vary; in homotopy (d) the end tangents are constant, but the closure is
only planar at the beginning and the end of the homotopy. To rule out
homotopies (c) and (d) (necessary for a well-defined writhe) we can require
the end tangents to be fixed and the closure to be planar throughout the
homotopy. It is possible that for a well-defined writhe further conditions
must be imposed.

\textbf{Euler-angle singularity \emph{vs.}\ the non-opposition condition.}
When the reference configuration is straight and end tangents
remain constant during the homotopy, the non-opposition condition
is equivalent to avoidance of the Euler-angle singularity at 
$\theta=\pi$. Although this is partly a coincidence, 
the two issues both stem from the topological properties of 
$S^2$.

The Euler-angle singularity results from the fact that $S^2$ is not 
homeomorphic to (any part of) $\R^2$. Any parametrization of
$S^2$ by a single cartesian coordinate system will therefore
have at least one singular point.
On the other hand, 
the non-opposition condition is necessary---in this article---for
the single-integral representation of writhe of Theorem~\ref{thm:fuller}.
In this representation the ambiguity of area `enclosed' by a curve
on $S^2$ is resolved by taking a perturbation approach. The non-opposition
condition is the realization of the unavoidable limits of this approach, and
therefore again stems from the topology of $S^2$.

As mentioned above, however, the non-opposition condition remains
an unsatisfactory element in the definition of open writhe. Perhaps a concept
of open writhe is possible that bypasses this condition.

\textbf{Open writhe is not rotation invariant.}
The definition of $\Ao$ depends on the choice of
the reference configuration. For a given reference configuration,
an open rod in $\Ao$ may not be freely rotated without
leaving $\Ao$. This is readily demonstrated by rotating
the reference configuration itself: after a rotation of $\pi$ about
an axis perpendicular to the plane of the reference curve the
non-opposition condition is violated at every 
point on the curve.

\begin{figure}[ht]
\centerline{\psfig{figure=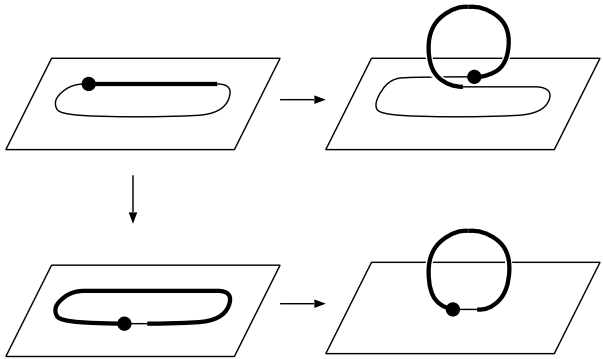,width=8cm}}
\caption{Two elements of $\Ao$ that differ only by
a rotation, but for which the writhe is different. The dot emphasizes the
difference in orientation.}
\label{fig:non_cont}
\end{figure}

This may lead to surprising results. In Figure~\ref{fig:non_cont}
two homotopies are shown. The first is a variation on homotopy~(b)
of Figure~\ref{fig:count_ex_genWrithe}, while in the second 
we lengthen the open-rod part and shorten the closure part.
In addition, we construct the homotopies such that the final
configurations are close, up to a rotation (emphasized by
the mark at one end of the open rod). In (a) the open writhe is close to
$1$, while in (b) it is close to $0$.

This remark also resolves an issue raised in Remark~\ref{rem:cont_conn}:
does the open writhe change continuously into the classical writhe for
closed rods, when an open rod is transformed into a closed
rod by lining up and connecting the ends? The answer is no---for the
resulting closed writhe would be rotation-invariant, contradicting
the remark above.

\section{Discussion}
\label{sec:disc}

The topological issues associated with large deformation discussed in this
paper are not of great concern in more traditional engineering applications.
As long as deformations are such that the integral in \pref{eq:link} remains
well-defined for a suitable choice of Euler angles (i.e., as long as the
angles stay away from the polar singularity) open 
link and end rotation are given
by \pref{eq:link}. However, in more modern applications of structural
mechanics, such as in molecular biology, large deformations occur more
routinely and more care is required. Indeed, Fuller's 1971 paper
\cite{fuller.71} was inspired by supercoiling DNA molecules. In this paper
the author also already introduces a (planar) closing curve in order to
compute the writhe of a simple (infinitely long) helix. Following the
pioneering work of Fuller an extensive literature has emerged on the
application of elastic rod theory to DNA supercoiling (\emph{cf}.~the survey
article by Schlick \cite{schlick.95}).

Open rods have become popular models for DNA molecules since by the early
1990s single-molecule experiments have become possible. First this involved
an applied force only \cite{smith.92}; later, once the molecule could be
prevented from swivelling at its (magnetically) loaded end, this involved
both an applied force and an applied moment \cite{strick.96}. Analytical
studies have addressed DNA in isolation \cite{benham.83,fain.97} as well as
in dilute solution using a statistical mechanics approach
\cite{marko.97,vologodskii.97,moroz.98,bouchiat.00}.

Benham \cite{benham.83} appears to have been be the first one to write
down an isoperimetric variational problem based on \pref{eq:link} in order
to find equilibrium configurations subject to constant link. He first uses
Fuller's result \pref{eq:fuller2} with $r_2$ taken to be a suitable closed
planar curve.
Then $Wr(r_2)=0$, and one obtains a single-integral formula for the
writhe of the curve in question $r_1$. When combined with the single integral
for the twist \pref{eq:twist_def}, this leads, via \pref{eq:ltw}, to a
single-integral expression for the link, which in a suitable Euler-angle
representation is given by \pref{eq:link}. Benham then observes that ``the
integral expressions for $Lk$, $Tw$ and $Wr$ may be constructed regardless of
whether the structure is closed''. Many workers have since followed Benham's
example to write link and writhe of an open rod as single integrals.

The implicit assumption in this approach is that there exists a continuous
deformation from $r_1$ to the planar curve $r_2$ which avoids opposition of
corresponding tangents. This may not be obvious,
since the non-opposition condition must be applied to the closed
combination of rod and suitable closure. In Section~\ref{sec:crit} 
we have given
examples of what can go wrong if the condition is not satisfied.
In addition, the example of Figure~\ref{fig:non_cont} shows that even
within the limits of the non-opposition condition the writhe is not
invariant under rotation, implying that  
this approach may lead to counterintuitive results.

Frequently, the end rotation needed in an energy analysis is simply assumed
to be equal to the link as given by \pref{eq:link}
(e.g., \cite{fain.97,moroz.98}). We have shown that end rotation can only
be defined in a consistent way within a class of homotopies of rods. Such
a class is constructed with the help of a closure. We define end rotation
independent of link and show the two to be equal within a suitable class of
allowed deformations. We should also remark that our closure is a rod
$(r,d_1)$
rather than just a centerline~$r$. Most authors initially only introduce a
closed centerline, which makes the writhe well-defined, and subsequently
assume the closure to be twistless if a link or end rotation is required.
We formalise this practice by explicitly specifying a $d_1$ for the entire
closed structure.

Knowing the precise restrictions on allowed deformations is important in
statistical mechanics studies. To obtain the correct averages one must
consider ensembles of admissible configurations. In numerical computations
this means that one must take care to simulate configurations (through a
Monte Carlo algorithm, a `growth' algorithm or otherwise) with the right
topological constraints. Specifically, one wants the configurations to be
unknotted and to have constant link (although it is good to remember that
DNA in its natural environment functions in the presence of topology-changing
enzymes). This means that one must forbid self-crossings of the
configuration as well as crossings of an (imaginary) closure. Mindful of
this, the authors in \cite{vologodskii.97} graft the ends of the molecule
to an external surface and run `sticks' out from the ends of the molecule
to infinity, thereby `virtually closing' the generated chain at infinity.
A similar construction is used in \cite{rossetto.xx}. In this latter work
knotted configurations are not eliminated, it being argued that the
molecular statistics is dominated by unknotted configurations.

In order to avoid the awkward non-opposition condition in Fuller's second
theorem there have been direct approaches via the double integral
\pref{eq:writhe_def} instead. In \cite{starostin,neukirch.preprint}
simple shapes are considered with planar closures for which the integral can
be evaluated explicitly. It is then shown that the closure gives a relative
contribution to the writhe which tends to zero as the length of the rod
tends to infinity. The double integral is also used in the numerical study
in \cite{vologodskii.97}, where it is shown that the contribution to $Wr$
from the interaction of the closure with the basic chain is of the order of
$1 \%$. Various numerical schemes for the computation of the writhe double
integral for a discretised curve are discussed and compared in
\cite{klenin.00}. Useful rigorous error bounds on numerically computed
values of $Wr$ based on polygonal (i.e., piece-wise linear) approximation
are given in \cite{cantarella.02}.

Our approach to a consistent definition of link, writhe and end rotation
is firmly based on the introduction of a closure. There have been various
formulations of writhe for open curves without the use of a closure. One
approach, especially taken in knot theory and in studies of self-avoiding
chains, is based on the characterisation of writhe as the average over all
planar projections of the sum of signed crossings \cite{fuller.71}
(e.g., \cite{orlandini.94,agarwal.02}). This approach does not require a
closure and can be applied to curves with arbitrary end tangents. It could
form the basis of an alternative extension of~\pref{eq:ltw} to open rods.
(Here we can remark that link also has an interpretation in terms of
crossing numbers, namely: the linking number of two curves is equal to
the number of all signed crossings of the curves in a regular planar
projection, i.e., one satisfying a transversality condition; see
\cite{rolfsen.76}.) The precise relation between the writhe of an open curve
obtained via these planar projections and the open writhe obtained by using
the tantrix area on $S^2$, or the double integral \pref{eq:writhe_def}, is
still an open problem.

If an exact writhe is not required and the fractional part modulo 2 is
sufficient then formula \pref{eq:fuller1} in terms of the area enclosed by
the tantrix on the unit sphere can be used. Rossetto \& Maggs
\cite{rossetto.xx} point out that this approach can also be used to
generalise the writhe to curves whose end tangents are not aligned and
therefore have open tantrices. By exploiting the connection between writhe
and a geometric phase they show that the canonical way to close the curve
is by means of a geodesic (great circle). This geodesic is unique as long
as the two end points are not antipodal. The fractional writhe is then
again given by the enclosed area. This prescription is used by Starostin
\cite{starostin} to derive results for the writhe of smooth as well as
polygonal curves. Cantarella \cite{cantarella.02} generalises Fuller's
second theorem, and with it the spherical area formula \pref{eq:fuller1},
to polygonal curves.

\appendix
\section{Proof of Theorem \ref{thm:wr-rot-invariance2}}
\label{sec:proofs}

Let $f:S^2\to\R$ be the function mentioned in the assertion.
Pick $v_0\in\Omega$ and let $\Omega_0$ be the connected component
of $\Omega$ containing $v_0$. 
Define the set 
\[
A = \{ v\in\Omega_0: f(v) = f(v_0)\}.
\]
The function $f$ is continuous on $\Omega_0$, implying that the set $A$ 
is relatively closed in $\Omega_0$. We will show below that
$f$ is constant on all open balls $B\subset \Omega_0$, implying that $A$ 
is also open. Since $A$ is non-empty it follows that $A = \Omega_0$ and
the Lemma is proved.

\medskip

For a given vector $\Om \in S^2$, let $R_{\phi}$ denote the 
rotation around $\Om$ through an angle~$\phi$. We fix
the direction of rotation in the following way: 
with respect to an orthonormal basis 
$(\Om, w, \Om \times w)$ for a suitable $w \in S^2$,
write $R_{\phi}$ as
\[
R_{\phi} = \left( 
\begin{array}{ccc}
1 & 0 & 0 \\
0 & \cos \phi & -\sin \phi \\
0 & \sin \phi & \cos \phi
\end{array}
\right).
\]
With this choice,
\begin{eqnarray}\label{eq:diff-rot}
\frac d{d\phi} R_{\phi}v|_{\phi = 0} = \Om \times v
\qquad \text{for any $v \in S^2$. }
\end{eqnarray}

Set $v_\phi = R_\phi v$.
Using equation~(\ref{eq:diff-rot}), we have
\begin{equation}
\label{eq:d1d2d3}
\frac d{d\phi} \frac {v_{\phi} \times t}{1 + v_{\phi}\cdot t} \cdot
\dot{t} = 
\frac{[(1 + v_{\phi}\cdot t)((\Om \times v_\phi)\times t) 
 - (v_{\phi} \times t)((\Om \times v_\phi)\cdot t)]}
{(1 + v_{\phi}\cdot t)^2}\cdot \dot t.
\end{equation}
Setting $\gamma=\modd{\omega\times v_\phi}$ we
introduce an orthonormal coordinate system 
\[
e_1=v_\phi, \quad e_2 = \gamma^{-1}\,\omega\times v_\phi,
\quad e_3 = \gamma^{-1}\,v_\phi\times(\omega\times v_\phi),
\]
and we write $t_1$, $t_2$, $t_3$ for the coordinates of $t$ with respect
to this basis; these are functions of the curve parameter $s$. 
The right-hand 
side of~\pref{eq:d1d2d3} becomes
\[
\gamma \, \frac {(1 + t_1)(t_3\dot t_1 - t_1 \dot t_3) -
t_2(t_2\dot t_3 - t_3 \dot t_2)}
{(1 + t_1)^2}.
\]
Using the equalities $t_1^2 + t_2^2 + t_3^2 = 1$ and 
$t_1\dot t_1 + t_2 \dot t_2 + t_3 \dot t_3 = 0$ this is seen to
be equal to 
\[
- \gamma \, \frac {d}{ds} \, \frac {t_3}{1+t_1}.
\]
Therefore
\[
\frac d{d\phi} \int \frac {v_{\phi} \times t}{1 + v_{\phi}\cdot t} \cdot
\dot{t}\, ds  = 
- \gamma \int \frac {d}{ds} \, \frac {t_3}{1+t_1} \,ds = 0.
\]
The last equality results from the assumption of aligned end tangents.
This proves the Theorem.

\section*{Acknowledgments}
This research was partly funded  under contract GR/R51698/01 by the EPSRC.
GH also thanks The Royal Society for continuing support.
The authors are grateful to Jason Cantarella for many interesting discussions,
and to Rubber Import Amsterdam BV for supplying us with ample 
experimental material.

\bibliography{refs}
\bibliographystyle{plain}

\end{document}